\documentclass[conference,a4paper]{IEEEtran}

\usepackage{cite}

\usepackage{graphicx,color,epsfig,rotating,subfigure}
\usepackage{amsfonts,amsmath,amssymb}
\usepackage{algorithm,algorithmic}
\usepackage{subfigure}

\long\def\comment#1{}

\newfont{\bbb}{msbm10 scaled 700}

\newfont{\bb}{msbm10 scaled 1100}

\newcommand{\PP}{\mbox{\bb P}}

\newcommand{\FF}{\mbox{\bb F}}

\newcommand{\ev}{{\bf e}}

\newcommand{\uv}{{\bf u}}
\newcommand{\wv}{{\bf w}}
\newcommand{\vv}{{\bf v}}
\newcommand{\xv}{{\bf x}}
\newcommand{\yv}{{\bf y}}

\newcommand{\zerov}{{\bf 0}}
\newcommand{\onev}{{\bf 1}}

\newcommand{\Cm}{{\bf C}}

\newcommand{\Em}{{\bf E}}

\newcommand{\Id}{{\bf I}}
\newcommand{\Jm}{{\bf J}}

\newcommand{\Qm}{{\bf Q}}

\newcommand{\Sm}{{\bf S}}

\newcommand{\Vm}{{\bf V}}

\newcommand{\Mc}{{\cal M}}

\newcommand{\Xc}{{\cal X}}
\newcommand{\Yc}{{\cal Y}}

\renewcommand{\det}{{\hbox{det}}}

\newcommand{\eqdef}{\stackrel{\Delta}{=}}

\newcommand{\transp}{{\sf T}}

\newtheorem{definition}{Definition}
\newtheorem{theorem}{Theorem}
\newtheorem{lemma}{Lemma}
\newtheorem{corollary}{Corollary}

\newtheorem{remark}{Remark}

\begin{document}

\sloppy

%% Paper Title
%% You can use linebreaks \\ within to get better formatting as
%% desired.
\title{Two-Unicast Two-Hop  Interference Network:\\ Finite-Field Model}

\author{
  \IEEEauthorblockN{Song-Nam~Hong}
  \IEEEauthorblockA{Dep. of Electrical Engineering\\
    University of Southern California\\
    CA, USA\\
    Email: songnamh@usc.edu}
  \and
  \IEEEauthorblockN{Giuseppe~Caire}
  \IEEEauthorblockA{Dep. of Electrical Engineering\\
    University of Southern California\\
    CA, USA\\
    Email: caire@usc.edu}
}
%

%% Create the title:
\maketitle

\begin{abstract}
In this paper we present a novel framework to convert the $K$-user multiple access channel (MAC) over $\FF_{p^m}$ into the $K$-user MAC over ground field $\FF_{p}$ with $m$ multiple inputs/outputs (MIMO). This framework makes it possible to develop coding schemes for MIMO channel as done in symbol extension for time-varying channel. Using aligned network diagonalization based on this framework, we show that the sum-rate of $(2m-1)\log{p}$ is achievable for a $2\times 2\times 2$ interference channel over $\FF_{p^m}$. We also provide some relation between field extension and symbol extension.
\end{abstract}

%%%%%%%%%%%%%%%%%%%%%%%%%%%%%%%%
\section{Introduction}

In recent years, significant progress has been made on the understanding of the theoretical limits of wireless communication networks. In \cite{Avestimehr}, the capacity of multiple multicast network (where every destination desires all messages) is approximated within a constant gap independent of SNR and of the realization of the channel coefficients. Also, for multiple flows over a single hop, new capacity approximations were obtained in the form of degrees of freedom (DoF), generalized degrees of freedom (GDoF), and $O(1)$ approximations \cite{Cadambe08,Gou09,Jafar10}. Yet, the study of multiple flows over multiple hops remains largely unsolved. The $2 \times 2 \times 2$ Gaussian interference channel (IC) has received much attention recently, being one of the fundamental building blocks to
characterize the DoFs of two-flows networks\cite{Shomorony}.
%One natural approach is to consider this model as a cascade of two ICs.
%In \cite{Simeone}, the authors apply the Han-Kobayashi scheme \cite{Han} for the first hop to split each message into private and common parts.
%Relays can cooperate using the shared information (i.e., common messages) for the second hop, in order to improve the data rates.
%This approach is known to be highly suboptimal at high signal-to-noise ratios (SNRs), since two-user IC can only achieve 1 DoF.
%In \cite{Cadambe}, Cadambe and Jafar show that $\frac{4}{3}$ DoF is achievable by viewing each hop as an X-channel.
%This is accomplished using the {\em interference alignment} scheme for each hop.
The optimal DoF was obtained in \cite{Gou} using {\em aligned interference neutralization},
which appropriately combines interference alignment and interference neutralization. Also, there was the recent extension to the $K \times K \times K$ Gaussian IC in \cite{Shomorony1}, achieving the optimal $K$ DoF using {\em aligned network diagonalization}.

In this paper we investigate interference networks over finite-field. This model can be meaningful in practical wireless communication systems, by the observation that the main bottleneck of a digital receiver is the Analog-to-Digital Conversion (ADC), which is power-hungry, and does not scale with Moore's law. Rather the number of bits per second produced by an ADC is roughly a constant that depends on the power consumption \cite{Walden}. Therefore, it makes sense to consider the ADC as part of channel, which may produce the finite-field model, as shown in \cite{Hong}.  Also, Compute-and-Forward (CoF) in  \cite{Nazer} enables to decode linear combinations of messages over finite-field at relays. By forwarding linear combinations, the overall end-to-end ``transfer function" between sources and destinations can be described by a system of linear equations over finite-field. Each destination can solve such equations to obtain desired messages as long as there exists a full-rank sub-system of equations involving the desired messages. In the setting of multiple flows (inference) over multiple hops, interference alignment (or neutralization and diagonalization) over finite-field is generally needed. However, current schemes developed for Gaussian channel may not be straightforwardly applicable for finite-field interference networks. For example, it is not so clear to apply the framework of real interference alignment \cite{Motahari} based on rational dimensions to finite-field interference networks.

\textbf{Our Contribution:} We show that the $K$-user multiple access channel (MAC) over $\FF_{p^m}$ is equivalent to the $K$-user MAC over $\FF_{p}$ with $m$ multiple inputs/outputs (MIMO).
In the transformed MIMO channel, the $m \times m$ channel matrices are represented by the powers of {\em companion matrix} of primitive element of $\FF_{p^m}$. This framework makes it possible to develop coding schemes for MIMO channel as done in symbol extension for time-varying channel. Next, we focus on a $2\times 2\times 2$ IC over $\FF_{p^m}$ and show that the sum-rate of $(2m-1)\log{p}$ is achievable by applying the concept of aligned network diagonalization to the transformed MIMO channel, under certain condition on channel coefficients. We also prove that this condition is satisfied with probability 1 if the channel coefficients are uniformly and independently drawn from non-zero elements of $\FF_{p^m}$ and either $m$ or $p$ goes to infinity. In addition, we consider the $2\times 2\times 2$ MIMO IC over $\FF_{p}$ and show that symbol extension (i.e., coding over multiple time slots) is needed for the aligned network diagonalization scheme. We characterize the required  symbol extension order (number of time slots over which coding takes place) that depends on the channel coefficients and is upper-bounded by the number of inputs/outputs.

%%%%%%%%%%%%%%%%%%%%%%%%%%%%%%%%%%%%%%%%%%%%%%%%%%%%%%%%%%
\section{MIMO Transform over ground field}\label{sec:MAC}
Throughout the paper, it is assume that  $\FF_{p^{m}}$ denotes a finite-field of order $p^m$, generated by a primitive polynomial   $\pi(x) \eqdef a_{0} + a_{1}x + \cdots + a_{m-1}x^{m-1} + x^{m}$.
The  elements of $\FF_{p^m}$ are given by the polynomial representation $\{b_{0}+b_{1}x +\cdots b_{m-1}x^{m-1}: b_{0},\ldots,b_{m-1} \in \FF_{p}\}$. Also, we can represent the elements of $\FF_{p^m}$ using primitive element $\alpha$ as $\{0=\alpha^{\infty},1=\alpha^{0},\alpha,\ldots,\alpha^{p^{m}-2}\}$. As usual, $\FF_{p^m}^{\star}$ denotes the multiplicative group of $\FF_{p^m}$, i.e., the set of non-zero elements of $\FF_{p^{m}}$.
%%%%%%%%%%%%%%%%%%%%%%%%%%
\begin{definition} The {\em companion matrix} of the polynomial $\pi(x)=a_{0} + a_{1}x + \cdots + a_{m-1}x^{m-1} + x^{m}$ is defined to be $m \times m$ matrix over $\FF_{p}$
\begin{equation*}
\Cm=\left[
    \begin{array}{ccccc}
      0 & 0 & \cdots& 0 & -a_{0} \\
      1 & 0 & \cdots &0 & -a_{1}\\
      0 & 1 & \cdots & 0 & -a_{2} \\
      \vdots & \vdots &\ddots &\vdots &\cdots\\
      0 & 0 & \cdots & 1 & -a_{m-1} \\
    \end{array}
  \right].
\end{equation*} \hfill $\lozenge$
\end{definition} Then,  $\Mc\eqdef\{\mbox{0}=\Cm^{\infty},\Id=\Cm^{0},\Cm,\ldots,\Cm^{p^{m}-2}\}$ forms a finite-field of order $p^m$. From \cite[Theorem 6]{Mac}, all finite fields of order $p^{m}$ are isomorphic~\footnote{Two fields $F,G$ are said to be {\em isomorphic} if there is a one-to-one mapping from $F$ onto $G$ which preserves addition and multiplication.}. Then, we have one-to-one mappings:
\begin{itemize}
\item Vector  representation (i.e., one-to-one mapping between polynomials and $m$-dimensional vectors over $\FF_{p}$):
\begin{equation}
\Phi(b_{0}+b_{1}x +\cdots +b_{m-1}x^{m-1})=[b_{0},\ldots,b_{m-1}]^{\transp}\label{eq:map1}.
\end{equation}
\item  Matrix representation (i.e., one-to-one mapping between elements of $\FF_{p^m}$ and matrices over $\FF_{p}$):
\begin{equation}
 \Psi(\alpha^{\ell}) = \Cm^{\ell}\label{eq:map2}.
\end{equation}
\end{itemize} With these mappings, we have:
\begin{lemma} For $\Xc_{k} \in \FF_{p^m}$, let $\Yc = \sum_{k=1}^{K} q_{k} \Xc_{k}$ for some coefficients $q_{k} \in \FF_{p^m}$. Also, set $\yv = \sum_{k=1}^{K}\Qm_{k}\xv_{k}$ where $\xv_{k} = \Phi(\Xc_{k}) \in \FF_{p}^{m}$ and  $\Qm_{k} = \Psi(q_{k}) \in \FF_{p}^{m \times m}$ for $k=1,\ldots,K$. Then, we have $\yv = \Phi(\Yc)$.\hfill\IEEEQED
\end{lemma}

The above lemma shows that the $K$-user {\em scalar} MAC over $\FF_{p^m}$ can be transformed into the $K$-user MIMO MAC over ground field $\FF_{p}$ where all nodes have $m$ multiple inputs/outputs.

\section{Two-Unicast Two-Hop IC over $\FF_{p^m}$}\label{sec:SISO}

We consider a $2\times 2\times 2$ IC over $\FF_{p^m}$ where all nodes have a single input/output. Notice that CoF framework produces a {\em noiseless} finite-field IC, while the symbol-by-symbol sampling (i.e., taking the ADC as part of channel) results in a finite-field IC with additive noise \cite{Hong}. In this paper we only consider a {\em noiseless} model by focusing  on interference management. In the first hop, the IC over $\FF_{p^m}$ is described by
\begin{equation}
\left[
  \begin{array}{c}
   \Yc_{1} \\
    \Yc_{2} \\
  \end{array}
\right] = \left[
            \begin{array}{cc}
              q_{11} & q_{12} \\
              q_{21} & q_{22} \\
            \end{array}
          \right]\left[
  \begin{array}{c}
    \Xc_{1} \\
   \Xc_{2} \\
  \end{array}
\right]\label{model:1-hop}
\end{equation} and also, in the second hop, the IC over $\FF_{p^m}$ is described by
\begin{equation}
\left[
  \begin{array}{c}
    \Yc_{3} \\
    \Yc_{4} \\
  \end{array}
\right] = \left[
            \begin{array}{cc}
              q_{33} & q_{34} \\
              q_{43} & q_{44} \\
            \end{array}
          \right]\left[
  \begin{array}{c}
   \Xc_{3} \\
    \Xc_{4} \\
  \end{array}
\right]\label{model:2-hop}
\end{equation} where $\Xc_{k} \in \FF_{p^m}, k=1,2,3,4$ and $\Yc_{\ell} \in \FF_{p^m}, \ell=1,2,3,4$. Here, the channel coefficients $q_{\ell k} \in \FF_{p^{m}}^{\star}$ are fixed and known to all nodes. Also, it is assumed that each hop has full-rank $2\times 2$ channel matrices over $\FF_{p^m}$.
%%%%%%%%%%%%%Figures%%%%%%%%%%%
\begin{figure}
\centerline{\includegraphics[width=7cm]{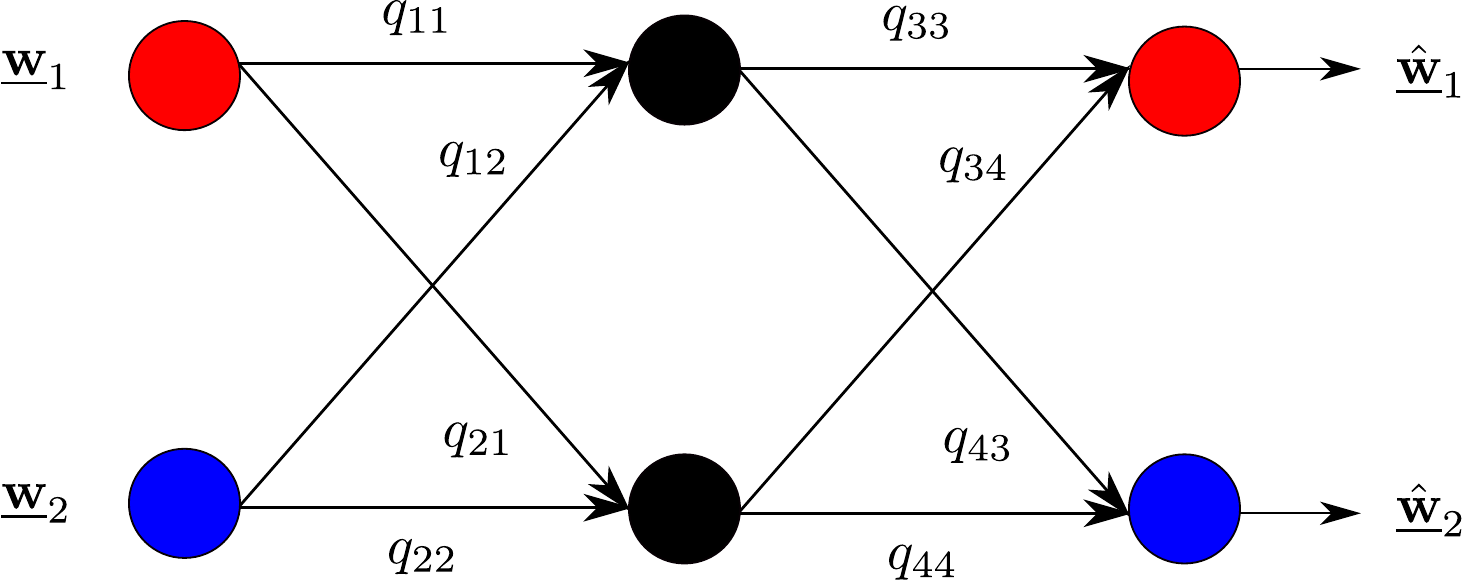}}
\caption{$2\times 2\times 2$ interference channel over $\FF_{p^m}$.}
\label{model222}
\end{figure}

%%%%%% Minimal Polynomial %%%%%%%%%%%%
\begin{definition}
The {\em minimal polynomial} over $\FF_{p}$ of $\beta \in \FF_{p^m}$ is the lowest degree monic polynomial $\mu(x)$ with coefficients from $\FF_{p}$ such that $\mu(\beta) = 0$. We denote the degree of polynomial $\mu(\beta)$ by  $\deg(\mu(\beta))$.\hfill $\lozenge$
\end{definition}

With this definition, we have:
%%%%%%%%%%%%%%%%%%%%%%%%%%%%
\begin{theorem}\label{thm} For $2\times 2\times 2$ IC over $\FF_{p^m}$,  the sum-rate of $(2m-1)\log{p}$ is achievable if $\deg(\mu(\gamma)) = m$ and $\deg(\mu(\gamma')) = m$ where
\begin{eqnarray}
\gamma = q_{11}^{-1}q_{12}q_{22}^{-1}q_{21}\mbox{ and }\gamma' = s_{11}^{-1}s_{12}s_{22}^{-1}s_{21}\label{eq:cond}
\end{eqnarray} and
\begin{equation*}
\left[
  \begin{array}{cc}
    s_{11} & s_{12} \\
    s_{21} & s_{22} \\
  \end{array}
\right] = \left[
            \begin{array}{cc}
              q_{33} & q_{34} \\
              q_{43} & q_{44} \\
            \end{array}
          \right]^{-1}.
\end{equation*}
\end{theorem}
\begin{IEEEproof}
See Section~\ref{sec:scheme}.
\end{IEEEproof}
%%%%%%%%%%%%%%%%%%%%%%%%%%%%%%%%%%%%%%%
Also, we derive a normalized achievable sum-rate with respect to interference-free channel capacity $m\log{p}$ when either $m$ or $p$ goes to infinity. This metric is analogous to degrees-of-freedom of Gaussian channels.

%%%%%%%%%%%%%%%%%%%%%%%%%%%%%%%%%%%%%%%%%%%%%%%%%%%%%%%%%%%%%%%%%%%%%%%%%%%%%%%%%%%%%%%%%%%%%%%%%%%%%%%%%%%%
\begin{corollary}\label{cor1} If the channel coefficients $q_{\ell k}$ are uniformly and independently drawn from $\FF_{p^m}^{\star}$, the following normalized sum-rates are achievable with probability $1$:
\begin{eqnarray}
d_{\rm{sum}}(p)&=&\lim_{m\rightarrow \infty}\frac{R_{\rm{sum}}(p,m)}{m\log{p}} =  2\label{eq:res1}\\
d_{\rm{sum}}(m)&=&\lim_{p \rightarrow \infty}\frac{R_{\rm{sum}}(p,m)}{m\log{p}} = \frac{2m-1}{m}\label{eq:res2}
\end{eqnarray}where $R_{\rm{sum}}(p,m)$ denotes the achievable sum-rate for given finite-field $\FF_{p^m}$.
\end{corollary}
\begin{IEEEproof} The proof consists of showing that the conditions in Theorem~\ref{thm} are satisfied with probability $1$ in the limits. Let $N(p,m)$ denote the number of {\em monic irreducible polynomials} of degree-$m$ over $\FF_{p}$. From \cite[Theorem 15]{Mac}, we have:
\begin{equation*}
N(p,m) = \frac{1}{m} \sum_{d|m} \nu(d)p^{m/d} \label{eq:irr}
\end{equation*} where $\nu(d)$ denotes the M\"{o}bius function, defined by
\begin{equation*}
\nu(d) = \left\{
           \begin{array}{ll}
             1 & \hbox{if } d=1 \\
             (-1)^{r}, & \hbox{if $d$ is the product of $r$ distinct primes} \\
             0, & \hbox{otherwise.}
           \end{array}
         \right.
\end{equation*} Notice that each degree-$m$ monic irreducible polynomial has $m$ distinct roots in $\FF_{p^m}$ and is a degree-$m$ minimal polynomial of such roots. Thus, we have $mN(p,m)$ distinct elements in $\FF_{p^m}$ with degree-$m$ minimal polynomial. Also, we can derive a simple lower-bound on  $mN(p,m)$ by setting $\nu(d) = -1$ for any  $d$ with $d|m, d > 1$:
\begin{equation*}
m N(p,m) \geq  p^{m} - \sum_{d|m, d>1}p^{m/d}.
\end{equation*} Using this bound and the fact that $\gamma$ defined in (\ref{eq:cond}), is uniformly distributed over $\FF_{p^m}^{\star}$, we can compute:
\begin{eqnarray*}
\PP(\{\deg(\mu(\gamma))=m\}) &=& \frac{mN(m,p)}{p^{m}-1}\\
&\geq& \frac{p^{m} - \sum_{d|m, d>1}p^{m/d}}{p^{m}}\\
&=& 1- \sum_{d|m, d>1}p^{m(1/d -1)}.
\end{eqnarray*} This probability goes to 1 if either $m$ or $p$ goes to infinity. With the same procedures, we can also prove that $\PP(\{\deg(\mu(\gamma'))=m\})$ goes to 1 if either $m$ or $p$ goes to infinity. This completes the proof.
\end{IEEEproof}

%%%%%%%%%%%%%%%%%%%%%%%%%%%%%%%%%%
\begin{remark}
We provide a brief comparison with the case of a $2\times 2\times 2$ IC over $\FF_{p}$ with time varying channel and $m$-symbol extension. Similarly to the case of the degree-$m$ extension field, the symbol extension also yields a MIMO IC where $m \times m$ channel matrices are the form of diagonal matrix with diagonal elements in $\FF_{p}^{\star}$. One may expect that the two MIMO channel models (namely, the one obtained by field extension and the other by symbol extension) are equivalent since they have about $p^m$  possible channel matrices and these matrices belong to a commutative algebra (products of such matrices do not depend on the order of the factors). For the symbol extension, the same achievable scheme of Section~\ref{sec:scheme} can be used under different feasibility conditions, namely, that the diagonal elements of the products of channel matrices (i.e., $\Qm=\Qm_{11}^{-1}\Qm_{12}\Qm_{22}^{-1}\Qm_{21}$ in (\ref{const1})) are distinct and non-zero \cite{Gou}. We can compute the probability that this condition is satisfied. For $p \rightarrow \infty$, the condition is satisfied with probability 1, as for the case of field extension. However, when $m \rightarrow \infty$ and $p$ is finite, this probability is strictly less than 1, while we have seen before that in the field extension the feasibility probability goes to 1 also in this case. This shows that symbol extension and field extension are generally not equivalent.
\hfill $\lozenge$
\end{remark}

%%%%%%%%%%%%%%%%%%%%%%%%%%%%%%%%%%%
\subsection{Proof of Theorem 1: Achievable scheme}\label{sec:scheme}

We prove Theorem~\ref{thm} using aligned network diagonalization, under the assumption that $\gamma$ and $\gamma'$ have degree-$m$ minimal polynomial. From Section~\ref{sec:MAC}, we can transform the $2\times 2\times 2$ scalar IC over $\FF_{p^m}$ in (\ref{model:1-hop}) and (\ref{model:2-hop}) into MIMO IC over $\FF_{p}$ with channel coefficients  $\Qm_{\ell k}= \Psi(q_{\ell k}) \in \FF_{p}^{m \times m}$. Notice that $\Qm_{\ell k}$ is always full rank over $\FF_{p}$. The proposed coding scheme is performed for the transformed MIMO channel and the one-to-one mapping $\Phi(\cdot)$ is used to transmit  coded messages via the channels.  In order to transmit $(2m-1)$ streams,  source $1$ sends $m$ independent messages
$\{w_{1,\ell} \in \FF_{p}: \ell=1,\ldots,m\}$ to destination 1 and source 2 sends $m-1$ independent messages $\{w_{2,\ell} \in \FF_{p}: \ell=1,\ldots,m-1\}$ to destination 2. For simplicity, we also use the vector representation of messages as $\wv_{1} = [w_{1,1},\ldots,w_{1,m}]^{\transp}$ and $\wv_{2} = [w_{2,1},\ldots,w_{2,m-1}]^{\transp}$.

\subsubsection{Encoding at the sources}

We let $\Vm_{1} = [\vv_{1,1},\ldots,\vv_{1,m}] \in \FF_{p}^{m \times m}$ and $\Vm_{2} = [\vv_{2,1},\ldots,\vv_{2,m-1}] \in \FF_{p}^{m \times m-1}$ denote the precoding matrices used at sources 1 and 2, respectively, chosen to satisfy the {\em alignment conditions}:
\begin{eqnarray}
\Qm_{11}\vv_{1,\ell+1} &=& \Qm_{12}\vv_{2,\ell}\nonumber\\
\Qm_{21}\vv_{1,\ell} &=& \Qm_{22}\vv_{2,\ell}\label{eq:alignment}
\end{eqnarray}for $\ell=1,\ldots,m-1$. For alignment, we use the construction method proposed in \cite{Gou}:
\begin{eqnarray}
\vv_{1,\ell+1} &=& (\Qm_{11}^{-1}\Qm_{12}\Qm_{22}^{-1}\Qm_{21})^{\ell}\vv_{1,1} \label{const1}\\
\vv_{2,\ell} &=& (\Qm_{22}^{-1}\Qm_{21}\Qm_{11}^{-1}\Qm_{12})^{\ell-1}\Qm_{22}^{-1}\Qm_{21}\vv_{1,1}\label{const2}
\end{eqnarray}for $\ell=1,\ldots,m-1$. Using $\Psi(\cdot)$ and $\gamma$ defined in (\ref{eq:cond}), the above constructions can be rewritten as
\begin{eqnarray}
\vv_{1,\ell+1} &=&  \Psi(q_{11}^{-1}q_{12}q_{22}^{-1}q_{21})^{\ell}\vv_{1,1} = \Psi(\gamma^{\ell})\vv_{1,1}\label{eq:const1}\\
\vv_{2,\ell} &=&\Psi(q_{22}^{-1}q_{21})\Psi(\gamma^{\ell-1})\vv_{1,1}\label{eq:const2}
\end{eqnarray}for $\ell=1,\ldots,m-1$.

\textbf{Encoding:}
\begin{itemize}
\item Source $k$ precodes its message over $\FF_{p}$ as $\xv_{k}=\Vm_{k}\wv_{k}$
and produces the channel input
\begin{equation}
\Xc_{k} = \Phi^{-1}(\xv_{k}) \in \FF_{p^m}, \;\;\; k=1,2.
\end{equation} Then, $\Xc_{1}$ and $\Xc_{2}$ are transmitted over channels.
\end{itemize}

\subsubsection{Relaying operations}

Relays decode linear combinations of source messages and forward the precoded linear combinations to destination.

\textbf{Decoding:}
\begin{itemize}
\item Relay 1 observes:
\begin{eqnarray*}
\Yc_{1} = q_{11}\Xc_{1} + q_{12}\Xc_{2} \in \FF_{p^m}
\end{eqnarray*} and maps the received signal onto ground field $\FF_{p}$:
\begin{eqnarray}
\Phi(\Yc_{1}) &=& \Qm_{11}\Phi(\Xc_{1})+ \Qm_{12}\Phi(\Xc_{2})\nonumber\\
&=& \Qm_{11}\Vm_{1}\wv_{1} + \Qm_{12}\Vm_{2}\wv_{2}\nonumber\\
&\stackrel{(a)}{=}&\Qm_{11}\Vm_{1} \underbrace{\left[
                     \begin{array}{c}
                       w_{1,1} \\
                       w_{1,2} + w_{2,1} \\
                       \vdots \\
                       w_{1,m} + w_{2,m-1} \\
                     \end{array}
                   \right]}_{\eqdef \uv_{1}} \label{received1}
\end{eqnarray} where (a) is due to the fact that precoding vectors satisfy the alignment conditions in (\ref{eq:alignment}). Since $\Vm_{1}$ is full-rank over $\FF_{p}$ by Lemma~\ref{lem:full}, relay 1 can decode $\uv_{1}$ (i.e., linear combinations of source messages).
\item Similarly, relay 2 observes the aligned signals over $\FF_{p}$:
\begin{eqnarray}
\Phi(\Yc_{2}) &=& \Qm_{21}\Phi(\Xc_{1}) + \Qm_{22}\Phi(\Xc_{2})\nonumber\\
&=& \Qm_{21}\Vm_{1}\wv_{1} + \Qm_{22}\Vm_{2}\wv_{2}\nonumber\\
&\stackrel{(a)}{=}&\Qm_{21}\Vm_{1}\underbrace{ \left[
                     \begin{array}{c}
                       w_{1,1} + w_{2,1} \\
                       \vdots \\
                       w_{1,m-1} + w_{2,m-1} \\
                       w_{1,m}\\
                     \end{array}
                   \right]}_{\eqdef \uv_{2}}\label{received2}
\end{eqnarray}where (a) is due to the fact that precoding vectors satisfy the alignment conditions in (\ref{eq:alignment}). Since $\Vm_{1}$ is full-rank over $\FF_{p}$ by Lemma~\ref{lem:full}, relay 2 can decode  $\uv_{2}$.
\end{itemize}

%%%%%%%%%%%%%%%%
\begin{lemma}\label{lem:full} Assume that $\deg(\mu(\gamma)) = m$. $\Vm_{1}$ has rank $m$ if we choose  $\vv_{1,1} = \Phi(1)$.
\end{lemma}
\begin{IEEEproof} Using $\vv_{1,1} = \Phi(1)$,  we have:
\begin{equation}
\Psi(\gamma^{\ell}) \vv_{1,1} = \Phi(\Psi^{-1}(\Psi(\gamma)^{\ell})\Phi^{-1}(\vv_{1,1})) = \Phi(\gamma^{\ell})\label{eq:form}.
\end{equation} From (\ref{eq:const1}) and  (\ref{eq:form}), the precoding matrix $\Vm_{1}$ can be written as
\begin{eqnarray*}
\Vm_{1} &=& [\vv_{1,1},\ldots,\vv_{1,m}] \\
&=&[\Phi(1), \Phi(\gamma), \Phi(\gamma^2),\ldots, \Phi(\gamma^{m-1})].\label{eq:V1}
\end{eqnarray*}  Since $\gamma$ is assumed to have degree-$m$ minimal polynomial, the following holds:
\begin{equation*}
b_{0} + b_{1}\gamma + \cdots + b_{m-1}\gamma^{m-1} \neq \zerov
\end{equation*} for any non-zero coefficients vector $(b_{0},\ldots,b_{m-1}) \in \FF_{p}^{m}$. Using this, we can prove that $\Vm_{1}$ has $m$ linearly independent columns:
\begin{eqnarray*}
&&b_{0}\Phi(1)+ b_{1}\Phi(\gamma) + \cdots + b_{m-1}\Phi(\gamma^{m-1}) \\
&=&\Phi(b_{0}) + \Phi(b_{1}\gamma) +  \cdots + \Phi(b_{m-1}\gamma^{m-1})\\
&=& \Phi(b_{0} + b_{1}\gamma + \cdots + b_{m-1}\gamma^{m-1})\neq \textbf{0}
\end{eqnarray*}  for any non-zero coefficients vector $(b_{0},\ldots,b_{m-1})  \in \FF_{p}^{m}$. This completes the proof.
\end{IEEEproof}

\textbf{Encoding:}

\begin{itemize}
\item Relay 1 precodes the decoded linear combinations as $\xv_{3} = \Sm_{11}\Vm_{3}\uv_{1}$  and produces the channel input
\begin{equation}
\Xc_{3}=\Phi^{-1}(\xv_{3}) \in \FF_{p^{m}}
\end{equation}
\item Likewise, relay 2 precodes the decoded linear combinations as $\xv_{4} = \Sm_{21}\Vm_{3}\uv_{2}$ and produces the channel input
\begin{equation}
\Xc_{4}=\Phi^{-1}(\xv_{4}) \in \FF_{p^{m}}
\end{equation} where
\begin{equation}
\Sm=\left[
       \begin{array}{cc}
         \Sm_{11} & \Sm_{12} \\
         \Sm_{21} & \Sm_{22} \\
       \end{array}
     \right] =\left[
       \begin{array}{cc}
         \Qm_{33} & \Qm_{34} \\
         \Qm_{43} & \Qm_{44} \\
       \end{array}
     \right]^{-1}\label{eq:S}
\end{equation}  and $\Vm_{3}$ are chosen to satisfy the alignment conditions in (\ref{eq:alignment}) with respect to $\Sm$:
\begin{eqnarray}
\vv_{3,\ell+1} &=& \Psi(\gamma'^{\ell})\vv_{3,1}\label{eq:const3}\\
\vv_{4,\ell} &=&\Psi(s_{22}^{-1}s_{21})\Psi(\gamma'^{\ell-1})\vv_{3,1}\label{eq:const4}
\end{eqnarray}for $\ell=1,\ldots,m-1$ where $s_{ij} = \Psi^{-1}(\Sm_{ij})$ and where $\gamma'$ is defined in (\ref{eq:cond}).
\end{itemize} From Lemma~\ref{lem:full}, we can immediately prove that $\Vm_{3}$ and $\Vm_{4}$ are full rank by choosing $\vv_{3,1}=\Phi(1)$ since $\deg(\mu(\gamma')) = m$. The other precoding vectors are completely determined by the (\ref{eq:const3}) and (\ref{eq:const4}).

From  (\ref{received1}) and (\ref{received2}), we can observe that the coefficients of the linear combinations only depend on alignment conditions, independent of channel coefficients. From this, we can produce the received signal for which the channel matrix is equal to the inverse of second-hop channel matrix. This is the key property to enable the network diagonalization. That is,
$\xv_{3}$ and $\xv_{4}$ are equal to received signals with channel coefficients $\Sm$:
\begin{eqnarray}
\left[
  \begin{array}{c}
   \xv_{3} \\
   \xv_{4} \\
  \end{array}
\right] &=&\left[
  \begin{array}{c}
    \Sm_{11}\Vm_{3}\uv_{1} \\
    \Sm_{21}\Vm_{3}\uv_{2} \\
  \end{array}
\right] = \left[
  \begin{array}{c}
    \Sm_{11}\Vm_{3} \wv_{1} + \Sm_{12}\Vm_{4}\wv_{2} \\
    \Sm_{21}\Vm_{3}\wv_{1} + \Sm_{22}\Vm_{4}\wv_{2} \\
  \end{array}
\right] \nonumber\\
&=& \left[
       \begin{array}{cc}
         \Qm_{33} & \Qm_{34} \\
         \Qm_{43} & \Qm_{44} \\
       \end{array}
     \right]^{-1}\left[
  \begin{array}{c}
    \Vm_{3}\wv_{1} \\
    \Vm_{3}\wv_{2} \\
  \end{array}
\right].\label{equiv}
\end{eqnarray}

\subsubsection{Decoding at the destinations}

Destinations 1 and 2 observe:
\begin{eqnarray*}
\left[
  \begin{array}{c}
    \Yc_{3} \\
    \Yc_{4} \\
  \end{array}
\right] = \left[
            \begin{array}{cc}
              q_{33} & q_{34} \\
              q_{43} & q_{44} \\
            \end{array}
          \right]\left[
  \begin{array}{c}
   \Xc_{3} \\
    \Xc_{4} \\
  \end{array}
\right].
\end{eqnarray*} By mapping the received signals onto ground field $\FF_{p}$, we can get:
\begin{eqnarray*}
\left[
  \begin{array}{c}
    \Phi(\Yc_{3}) \\
    \Phi(\Yc_{4}) \\
  \end{array}
\right]
&=&  \left[
            \begin{array}{cc}
              \Qm_{33} & \Qm_{34} \\
              \Qm_{43} & \Qm_{44} \\
            \end{array}
          \right] \left[
                              \begin{array}{c}
                              \Sm_{11}\Vm_{3}\uv_{1} \\
                                \Sm_{21}\Vm_{3}\uv_{2} \\
                              \end{array}
                            \right]\\
                            &\stackrel{(a)}{=}& \left[
                              \begin{array}{c}
                              \Vm_{3}\wv_{1} \\
                                \Vm_{4}\wv_{2} \\
                              \end{array}
                            \right]
\end{eqnarray*}  where (a) is due to the precoding at relays to satisfy (\ref{equiv}). This shows that destination 1 can decode $\wv_{1}$ using $\Vm_{3}^{-1}\Phi(\Yc_{3})$ and destination 2 can decode $\wv_{2}$ using $\Vm_{4}^{-1}\Phi(\Yc_{4})$. This completes the proof of Theorem~\ref{thm}.

%%%%%%%%%%%%%%%%%%%%%%%%%%%%%%%%%%%%%%%%%%%%%%%%%
%%%%%%%%%%%%%%%%%%%%%%%%%%%%%%%%%%%%%%%%%%%%%%%%%
\section{Two-Unicast Two-Hop MIMO IC over $\FF_{p}$}\label{sec:MIMO}

We consider a $2\times 2\times 2$ MIMO IC over $\FF_{p}$ where all nodes have $m$ multiple inputs/outputs. Here, the $m\times m$ channel matrices are denoted by
$\Qm_{\ell k} \in \FF_{p}^{m\times m}$. Notice that they are neither diagonal matrices nor in the form of powers of companion matrix, and do not commute. Therefore, it is not possible to apply straightforwardly the same approach developed before. Instead, we have to resort to symbol extension by going to an extension field in order to obtain aligned network diagonalization.

From (\ref{eq:const1}), we can define the precoding matrix $\Vm_{1}$ to satisfy the alignment conditions as function of $\vv_{1,1}$:
\begin{equation}
\Vm_{1} = [\vv_{1,1},\Qm\vv_{1,1}\ldots,\Qm^{m-1}\vv_{1,1}] \
\end{equation} where $\Qm = \Qm_{11}^{-1}\Qm_{12}\Qm_{22}^{-1}\Qm_{21}$. We cannot use the result in Section~\ref{sec:scheme} since $\Qm$ is not mapped onto the element of $\FF_{p^m}$. For the time being, we assume that $\Qm$ has $m$ distinct eigenvalues. Following \cite{Gou,Hong-J}, we can prove that $\Vm_{1}$ is full rank if we choose $\vv_{1,1} = \Em\onev$ where  $\Em$ consists of $m$ linearly independent eigenvectors of $\Qm$. In case of complex-valued Gaussian channel, we can always find $m$ distinct eigenvalues in the given complex field. However, in the finite field $\FF_{p}$, some eigenvalues of $\Qm$ may not exist in the ground field $\FF_{p}$, depending on characteristic polynomial of $\Qm$ (denoted by $C(\lambda)$). Suppose that this polynomial is factored in the following way:
\begin{equation}
C(\lambda)=\prod_{i}\pi_{i}(\lambda)
\end{equation} where $\deg(\pi_{i}(\lambda)) \geq \deg(\pi_{j}(\lambda))$ if $i \leq j$. If $\deg(\pi_{1}(\lambda)) = r > 1$ then some eigenvalues of $\Qm$ do not exist in $\FF_{p}$. Also, we can see that $\pi_{1}(\lambda)$ is a degree-$r$ irreducible polynomial over $\FF_{p}$. Thus,  $L=\FF_{p}[\lambda]/\pi_{1}(\lambda)$ generates an extension field of $\FF_{p}$ with order $p^{r}$ and is isomorphic to $\FF_{p^r}$. We can notice that $r$ is the minimum order for which the corresponding extension field contains the roots of $\pi_{1}(\lambda)$. Since $\deg(\pi_{i}(\lambda)) \leq r$ for $i>1$, we are able to find all roots of $C(\lambda)$ in $\FF_{p^r}$.  In short, $\FF_{p^r}$ is the  {\em splitting field}~\footnote{A splitting field of a polynomial with coefficients in a field is a smallest field extension of that field over which the polynomial splits into linear factors.} of $C(\lambda)$. Assume that $\Qm$ has $m$ distinct eigenvalues $\{\lambda_{i} \in \FF_{p^r}: i=1,\ldots,m\}$ and corresponding eigenvectors $\{\ev_{i} \in \FF_{p^{r}}^{m}: i=1,\ldots,m\}$. Since $\Qm$ is diagonalizable, we have
$\Qm = \Em\Lambda\Em^{-1}$ where $\Em$ has $\ev_{i}$ as its the $i$-th column and $\Lambda$ has $\lambda_{i}$ as its $i$-th diagonal element.
Then, we choose  $\vv_{1,1} = \Em\onev \in \FF_{p^r}^{m}$. Following \cite{Hong-J}, we can show that $\Vm_{1}$ is full rank over $\FF_{p^r}$ as follows. Since  $\Qm=\Em\Lambda\Em^{-1}$ and $\vv_{1,1} = \Em\onev$, we have:
\begin{eqnarray}
\Vm_{1} = \Em\underbrace{\left[
         \begin{array}{cccc}
           1 & \lambda_{1} & \cdots & \lambda_{1}^{m-1} \\
           \vdots & \vdots & \ddots & \vdots \\
           1 & \lambda_{m} & \cdots & \lambda_{m}^{m-1} \\
         \end{array}
       \right]}_{\eqdef \Jm}.
\end{eqnarray} Since $\Jm$ is a Vandermonde matrix, the determinant of $\Vm_{1}$ is computed by
\begin{eqnarray*}
\det(\Vm_{1}) &=& \det(\Em)\det(\Jm)\\
&=&\det(\Em)\prod_{1\leq i < j\leq m}(\lambda_{j} - \lambda_{i}) \neq 0.
\end{eqnarray*} Therefore, $\Vm_{1}$ is full rank.

Next, we  present our coding scheme over the $r$-symbol extension (i.e., over $r$ time slots).

\textbf{Encoding at the sources:}
\begin{itemize}
\item Source 1 precodes its message $\wv_{1} \in \FF_{p^r}^{m}$ using precoding matrix $\Vm_{1} \in \FF_{p^r}^{m\times m}$:
\begin{equation*}
\xv_{1} = \Vm_{1}\wv_{1} \in \FF_{p^r}^{m}
\end{equation*}  and transmits the $t$-th column of $\Phi^{\transp}(\xv_{1}) \in \FF_{p}^{m \times r}$ at time slot $t$ for $t=1,\ldots,r$ where $\Phi^{\transp}: \FF_{p^r} \rightarrow [\FF_{p},\ldots,\FF_{p}]$ (notice that differently from (\ref{eq:map1}), it maps the elements of $\FF_{p^r}$ to the $r$-dimensional row vectors).
\item Similarly, source 2 precodes its message $\wv_{2} \in \FF_{p^r}^{m-1}$ using precoding matrix $\Vm_{2} \in \FF_{p^r}^{m\times m-1}$:
\begin{equation*}
\xv_{2} = \Vm_{2}\wv_{2} \in \FF_{p^r}^{m}
\end{equation*}  and transmits the $t$-th column of $\Phi^{\transp}(\xv_{2}) \in \FF_{p}^{m \times r}$ at time slot $t$ for $t=1,\ldots,r$.
\end{itemize}

\textbf{Decoding at the relays:}
\begin{itemize}
\item Relay 1 observes:
\begin{eqnarray*}
\Phi^{\transp}(\yv_{1}) &=& \Qm_{11}\Phi^{\transp}(\xv_{1}) + \Qm_{12}\Phi^{\transp}(\xv_{2}) \in \FF_{p}^{m \times r}.
\end{eqnarray*} By mapping the received signal onto the element of $\FF_{p^{r}}$, we have:
\begin{eqnarray*}
\yv_{1} &=& \Qm_{11}\Vm_{1}\wv_{1} + \Qm_{12}\Vm_{2}\wv_{2}\\
&=& \Qm_{11}\Vm_{1}\uv_{1}
\end{eqnarray*} where the last step is due to the fact that precoding vectors satisfy the alignment conditions in (\ref{eq:alignment}).
\item Similarly, relay 2 observes the aligned signal:
\begin{eqnarray*}
\yv_{2} &=& \Qm_{21}\Vm_{1}\uv_{2}.
\end{eqnarray*}
\end{itemize} At this point, we can follow Section~\ref{sec:scheme}. In this case, we can achieve the sum-rate of $(2m-1)\log{p^r}$  during $r$ time slots. Therefore, we can achieve the sum-rate of $(2m-1)\log{p}$ per time slot.

\begin{remark}The number of required symbol extensions $r \leq m$ depends on the channel coefficients. In general, we can always use the $m$-symbol extension to use the aligned network diagonalization, regardless of channel coefficients. In this way, the coding block length (symbol extension order) depends only on the number of inputs/outputs at each node, and it is independent of the channel coefficients.
\hfill $\lozenge$
\end{remark}

\section*{Acknowledgment}

This work was supported by NSF Grant CCF 1161801.

%%%%%%%%%%%%%%%%%%%%%%%%%%%%%%%%%%%%%%%%%%%


\begin{thebibliography}{1}




\bibitem{Avestimehr} S. Avestimehr, S. Diggavi, and D. Tse, ``Wireless network information flow: A deterministic approach," {\em IEEE Transactions on Information Theory,} vol. 57, pp. 1872-1905, Apr. 2011.

\bibitem{Cadambe08} V. Cadambe and S. Jafar, ``Interference alignment and the degrees of freedom of the K user interference channel," {\em IEEE Transactions on Information Theory,} vol. 54, pp. 3425-3441, Aug. 2008.

\bibitem{Gou09} T. Gou and S. A. Jafar, ``Capacity of a class of symmetric SIMO Gaussian interference channels within O(1)," in {\em Proceedings of IEEE International Symposium on Information Theory (ISIT),} Seoul, Korea, Jun-Jul. 2009.

\bibitem{Jafar10} S. A. Jafar and S. Vishwanath, ``Generalized Degrees of Freedom of the Symmetric Gaussian K User Interference Channel," {\em IEEE Transactions on Information Theory,} vol. 56, pp. 3297-3303, Jul. 2010.

\bibitem{Shomorony} I. Shomorony and S. Avestimehr, ``Two-Unicast Wireless Networks: Characterizing the Degrees-of-Freedom," {\em IEEE Transactions on Information Theory,} vol. 59, pp. 353-383, Jan. 2013.

%\bibitem{Simeone} O. Simeone, O. Somekh, Y. Bar-Ness, H. V. Poor, and S. Shamai, ``Capacity of Linear Two-Hop Mesh Networks with Rate Splitting, Decode-and-Forward Relaying and Cooperation," in {\em Proceedings of 45th Annual Allerton Conference on Communication, Control, and Computing,} Monticello, Illinois, Sept. 26-28, 2007.

%\bibitem{Han} T. Han and K. Kobayashi, ``A New Achievable Rate Region for the Interference Channel," {\em IEEE Transactions on Information Theory,} vol. IT-27, pp. 49-60, Jan. 1981.

%\bibitem{Cadambe} V. R. Cadambe and S. A. Jafar, ``Interference Alignment and Degrees of Freedom of Wireless X Networks," {\em IEEE Transactions on Information Theory,} vol. 55, pp. 3893-3908, Sept. 2009.

\bibitem{Gou} T. Gou, S. A. Jafar, S.-W. Jeon, S.-Y. Chung, ``Interference Alignment Neutralization and the Degrees of Freedom of the $2\times 2\times 2$ Interference Channel, {\em IEEE Transactions on
    Information Theory,} vol. 58, pp. 4381-4395, July, 2012.

\bibitem{Shomorony1} I. Shomorony and S. Avestimehr, ``Degrees of Freedom of Two-Hop Wireless Networks: ``Everyone Gets the Entire Cake"," {\em To appear in proceedings of 2012 Allerton Conference}.


\bibitem{Walden} R. Walden, ``Analog-to-Digital Converter Survey and Analysis," {\em IEEE Journal on Selected Areas in Communications,} vol. 17. pp. 539-550, Apr. 1999.


\bibitem{Motahari} A. S. Motahari, S. O. Gharan, M. A. Maddah-Ali, and A. K. Khandani, ``Real Interference Alignment: Exploiting the Potential of Single Antenna Systems," {\em Submitted to IEEE Transactions on Information Theory} 2009.

%\bibitem{Zhan} J. Zhan, B. Nazer, U. Erez, and M. Gastpar, ``Integer-Forcing Linear Receivers," {\em submitted to IEEE Transactions on Information Theory}.

\bibitem{Hong} S. Hong and G. Caire, ``Compute-and-Forward Strategy for Cooperative Distributed Antenna Systems," {\em submitted to IEEE Transactions on Information Theory} 2012.


\bibitem{Nazer} B. Nazer and M. Gastpar, ``Compute-and-Forward: Harnessing Interference through Structured Codes," {\em IEEE Transactions on Information Theory,} vol. 57, pp. 6463-6486, Oct. 2011.

\bibitem{Mac} F. J. MacWilliams and N. J. A. Sloane, ``The Theory of Error-Correcting Codes," Bell Laboratories Murray Hill.

%\bibitem{Erez2004} U. Erez and R. Zamir, ``Achieving $\frac{1}{2}\log(1+\SNR)$ on the AWGN channel with lattice encoding and decoding," {\em IEEE Transactions on Information Theory,} vol. 50, pp. 2293-2314, Oct. 2004.
%
%\bibitem{Niesen1} U. Niesen and P. Whiting, ``The degrees-of-freedom of compute-and-forward," {\em IEEE Transactions on Information Theory,} vol. 59, pp. 5214-5232, Aug. 2012.

%\bibitem{Hong222} S.-N. Hong and G. Caire, ``Structured Lattice Codes for $2\times 2\times 2$ MIMO Interference Channel," {\em Submitted to IEEE International Symposium on Information Theory}.

\bibitem{Hong-J} S.-N. Hong and G. Caire, ``Structured Lattice Codes for Some Two-User Gaussian Networks with Cognition, Coordination, and Two-Hops," {\em Submitted to IEEE Transactions on Information Theory,} Apr. 2013.
\end{thebibliography}
\end{document}